\documentclass[a4paper,12pt]{article}

\usepackage{amsmath}
\usepackage[psamsfonts]{amssymb}
\usepackage{rsfs}
\usepackage{bm}

\usepackage{cite}
\usepackage[dvips]{graphicx}
\usepackage{color}

\begin{document} 

\begin{titlepage}

\baselineskip 10pt
\hrule 
\vskip 5pt
\leftline{}
\leftline{Chiba Univ. Preprint
          \hfill   \small \hbox{\bf CHIBA-EP-151}}
\leftline{\hfill   \small \hbox{hep-th/0504107}}
\leftline{\hfill   \small \hbox{December 2005}}
\vskip 5pt
\baselineskip 14pt
\hrule 
\vskip 1.0cm
\centerline{\Large\bf 
Yang-Mills Theory
}
\vskip 0.4cm
\centerline{\Large\bf  
Constructed from 
}
\vskip 0.4cm
\centerline{\Large\bf  
Cho--Faddeev--Niemi Decomposition
}

\vskip 0.5cm

\centerline{{\bf 
Kei-Ichi Kondo,$^{\dagger,\ddagger,{1}}$  
Takeharu Murakami$^{\ddagger,{2}}$
and 
Toru Shinohara$^{\ddagger,{3}}$
}}  
\vskip 0.5cm
\centerline{\it
${}^{\dagger}$Department of Physics, Faculty of Science, 
Chiba University, Chiba 263-8522, Japan
}
\vskip 0.3cm
\centerline{\it
${}^{\ddagger}$Graduate School of Science and Technology, 
Chiba University, Chiba 263-8522, Japan
}
\vskip 1cm

\begin{abstract}
We give a new way of looking at the Cho--Faddeev--Niemi (CFN) decomposition of the Yang-Mills theory 
to answer how the enlarged local gauge symmetry respected by the CFN variables is restricted to obtain another Yang-Mills theory with the same local and global gauge symmetries as the original Yang-Mills theory. 
This may shed new light on the fundamental issue of the discrepancy between two theories for independent degrees of freedom and the role of the Maximal Abelian gauge in Yang-Mills theory.  
As a byproduct, this consideration gives new insight into the meaning of the gauge invariance and the observables, e.g., a gauge-invariant mass term and   vacuum condensates of mass dimension two. 
We point out the implications for the Skyrme--Faddeev model.

\end{abstract}


\hrule  
\vskip 0.1cm
${}^1$ 
  E-mail:  {\tt kondok@faculty.chiba-u.jp}
  
${}^2$ 
  E-mail:  {\tt tom@cuphd.nd.chiba-u.ac.jp}
  
${}^3$ 
  E-mail:  {\tt sinohara@cuphd.nd.chiba-u.ac.jp}

\par 
\par\noindent


\vskip 0.5cm

\newpage
\pagenumbering{roman}




\end{titlepage}


\pagenumbering{arabic}

\baselineskip 14pt
\section{Introduction}
\setcounter{equation}{0}

In understanding a non-perturbative feature of quantum field theory, it is quite important to extract the dynamical (local) or topological (global) degrees of freedom which are most relevant to the physics in question. 
For example, it is widely accepted that the magnetic monopole is responsible for quark confinement and the Yang-Mills instanton for chiral symmetry breaking.

From this viewpoint, the decomposition or the change of variables of the Yang-Mills gauge field proposed by Cho \cite{Cho80}, Faddeev and Niemi \cite{FN99} 
(CFN) is very interesting, since it enables us to extract explicitly certain types of topological configurations in Yang-Mills theory, especially, the magnetic monopole of Wu--Yang type and a multi-instanton of Witten type (see e.g., \cite{TTF00} for the mutual dependence). 
A characteristic feature of the {\it CFN decomposition} for the SU(2) Yang-Mills gauge field $\mathscr{A}_\mu(x)$ is to introduce a three-dimensional unit vector field $\bm{n}(x)=(n^1(x), n^2(x), n^3(x))$  satisfying $\bm{n}(x) \cdot \bm{n}(x) :=n^A(x) n^A(x) = 1$; that is, there are two extra degrees of freedom, in addition to the two vector fields: $C_\mu(x)\bm{n}(x)$, parallel to $\bm{n}(x)$, and $\mathbb{X}_\mu(x)$, perpendicular to $\bm{n}(x)$ [and hence $\bm{n}(x) \cdot \mathbb{X}_\mu(x)=0$]. 
The $\bm{n}$ field represents the color direction (in a gauge invariant way) and plays the distinguished role of expressing the topological configurations mentioned above. 
An on-shell decomposition of the gauge field is also given in Faddeev--Niemi \cite{FN99}, in agreement with the on-shell counting of the degrees of freedom in the original Yang-Mills field.  This is very useful for the purpose of finding the classical solution of the equation of motion.   
The CFN decomposition is further developed in \cite{Shabanov99a,Shabanov99b,Gies01}.

The CFN decomposition has enormous potential which has enabled us to discover   novel non-perturbative features previously overlooked  in  Yang-Mills theory.
For example, the Skyrme--Faddeev model \cite{FN97} describing a glueball as a knot soliton solution can be deduced from the Yang-Mills theory by way of the  CFN decomposition of the Yang-Mills theory \cite{LN99,Gies01,BCK01,FN02,Cho03,Kondo04,KKMSS05,KKMSSI05} (see, e.g., \cite{Hirayama} for  exact solutions). 
Moreover, the stability recovery of the Savvidy vacuum \cite{Savvidy77} through the elimination of a tachyon mode \cite{NO78} has been shown \cite{Cho03,Kondo04,KKMSS05} by using the CFN decomposition. (See \cite{Cho03} for   treatment of a massless gluon and \cite{Kondo04,KKMSS05} for treatment of a massive gluon caused by novel magnetic condensation.)  Also,  numerical simulations on a lattice can be performed based on the CFN decomposition  \cite{KKMSS05,KKMSSI05}.

As mentioned above, however, the CFN change of variables introduces two extra degrees of freedom. 
At the level of  classical Yang-Mills theory, this does not cause any subtle problems. 
However, a question arises when we consider the quantization of the Yang-Mills theory written in terms of the CFN variables, which we call the {\it CFN-Yang-Mills theory} (the extended Yang-Mills theory), because the reparametrization seems to increase the number of dynamical degrees of freedom in the Yang-Mills theory.   
How do we deal with the two extra degrees of freedom introduced by $\bm{n}$?
This question has caused much controversy concerning the treatment and  interpretation of $\bm{n}$.  

A partial answer was given by Shabanov \cite{Shabanov99b}:  
 To obtain the same degrees of freedom as in the original Yang-Mills theory,  two constraint conditions, $\bm{\chi}(x)=0$, must be imposed, e.g., using the Lie-algebra valued functional $\bm{\chi}(x):=\chi^A(x)T^A$ subject to $\bm{n}(x) \cdot \bm{\chi}(x)=0$.  
In fact, the integration measure for the CFN variables in the framework of the functional integration method  has been constructed so that it is completely equivalent to the standard integration measure of the original Yang-Mills theory, provided that the two conditions $\bm{\chi}(x)=0$ are imposed on the measure in accordance with the above viewpoint.

 Some subtle aspects of this issue have been clarified and resolved by Cho and his collaborators \cite{BCK01}. 
They pointed out that it is meaningless to eliminate the two extra degrees of freedom created by $\bm{n}$ by adding two extra constraints. 
The constraint should be regarded as a gauge fixing condition or a consistency condition.
Moreover,  the topological field $\bm{n}$ becomes dynamical with the gauge fixing, increasing the number of dynamical degrees of freedom. 
They argued that the extended Yang-Mills theory is modified in a subtle but important way and they have proposed three quantization schemes based on different choices of the decomposition (or reparametrization), but neither  their uniqueness nor equivalence has been demonstrated.  
This clearly shows that there  remain the unresolved questions concerning the CFN decompositions. 
The lack of a complete understanding of the CFN decomposition is merely an  obstacle to utilizing this machinery.

The purpose of this paper is to reconsider the meaning of the CFN decomposition from a different viewpoint which could give a thorough and unambiguous understanding of the CFN-Yang-Mills theory. 
In particular, we explicitly specify  {\it the gauge group for the enlarged gauge symmetry of the CFN-Yang--Mills theory}, rather than merely counting the degrees of freedom. 
Then we can answer the following questions. 
\begin{enumerate}
\item
What gauge symmetry does the CFN-Yang-Mills theory possess?
Which part of the enlarged gauge symmetry of the CFN-Yang--Mills theory must be constrained to reproduce the gauge theory with the same gauge symmetry as the original Yang-Mills theory?
How do we fix the enlarged gauge symmetry to meet this requirement?
\item
What are the gauge invariant observables written in terms of the CFN variables?   

\item
How do we define the CFN decomposition on a lattice, and how do we perform numerical simulations of the lattice CFN-Yang--Mills theory \cite{KKMSS05,KKMSSI05}? 

\item
What is the correct form of the Faddeev--Popov ghost term in the BRST quantization in the continuum formulation \cite{KMS05b}?

\end{enumerate}
These are advantages of our way of interpreting the CFN variables from a new viewpoint. 
In this paper, we discuss the first two issues, while other issues are reported elsewhere \cite{KKMSS05,KKMSSI05,KMS05b}. 



\section{Yang-Mills theory in the CFN decomposition}
\setcounter{equation}{0}


\subsection{Local gauge symmetry in terms of the CFN variables}


The Cho--Faddeev-Niemi (CFN) decomposition (or change of variables) of the original Yang-Mills gauge field $\mathscr{A}_\mu(x)$ is performed as follows. 
We restrict our consideration to the gauge group $G=SU(2)$.
First of all, we introduce a unit vector field $\bm{n}(x)$ as
\begin{align}
  \bm{n}(x) \cdot \bm{n}(x) := n^A(x) n^A(x) = 1 \quad (A=1,2,3) . 
  \label{nn=1}
\end{align}
Then the off-shell CFN decomposition is written in the form 
\begin{equation}
\mathscr{A}_\mu(x)
 =c_\mu(x) \bm{n}(x)
  +g^{-1}\partial_\mu \bm{n}(x)\times \bm{n}(x)
  +\mathbb X_\mu(x) ,
\label{CFN}
\end{equation}
where $\mathbb{X}_\mu(x)$ is perpendicular to $\bm{n}$:
\begin{align}
  \bm{n}(x) \cdot \mathbb{X}_\mu(x) = 0 .
  \label{nX=0}
\end{align}
The first term on the right-hand side of (\ref{CFN}) is denoted by 
$
  \mathbb{C}_\mu(x) := c_\mu(x){\bm n}(x) .
$
It 
is parallel to $\bm{n}(x)$ and is called the restricted potential.
The second term is denoted 
$
  \mathbb{B}_\mu(x) := g^{-1}\partial_\mu{\bm n}(x)\times{\bm n}(x) .
$
It 
is perpendicular to $\bm{n}(x)$ and is called the magnetic potential. 
For later convenience, we define the sum of $\mathbb{C}_\mu(x)$ and  $\mathbb{B}_\mu(x)$:
\begin{align}
    \mathbb{V}_\mu(x)  :=  \mathbb{C}_\mu(x)  +  \mathbb{B}_\mu(x)
    = c_\mu(x) \bm{n}(x)
  +g^{-1}\partial_\mu \bm{n}(x)\times \bm{n}(x) .
\end{align}
The form of $\mathbb{V}_\mu(x)$ is determined by the requirement that the field $\bm{n}(x)$ be a covariant constant in the background field $\mathbb{V}(x)$:
\begin{align}
  D_\mu[\mathbb{V}] \bm{n} := \partial_\mu \bm{n} + g\mathbb{V}_\mu \times \bm{n}  = 0 .
  \label{D[V]n=0}
\end{align}
Here,  only the perpendicular part, $\mathbb{B}_\mu$, is uniquely determined, while the parallel component $c_\mu$ is not determined uniquely \cite{Manton77}. (In fact, any four-vector is allowed.)

As pointed out in \cite{Shabanov99b}, the restricted potential $c_\mu$ and the gauge covariant potential $\mathbb{X}_\mu$ are specified by $\bm{n}$ and $\mathscr{A}_\mu$ as 
\begin{align}
c_\mu(x)
 &={\bm n}(x)\cdot\mathscr{A}_\mu(x), 
\quad 
\mathbb X_\mu(x)
  =g^{-1}{\bm n}(x)\times D_\mu[\mathscr{A}]{\bm n}(x) .
\label{def:X}
\end{align}
The first equation is obtained from (\ref{nX=0}) and  (\ref{nn=1}), which yield  $\bm{n}(x) \cdot \partial_\mu \bm{n}(x)=0$, while  the second equation is obtained by making use of the fact (\ref{D[V]n=0}), 
which yields 
\begin{align}
  D_\mu[\mathscr{A}] \bm{n} 
  := \partial_\mu \bm{n} +  g\mathscr{A}_\mu \times \bm{n} 
  = D_\mu[\mathbb{V}] \bm{n} + g\mathbb{X}_\mu \times \bm{n} 
= g\mathbb{X}_\mu \times \bm{n} .
\label{XnA}
\end{align}
More explicitly, the Yang-Mills gauge field $\mathscr{A}_\mu(x)$ can be cast into the equivalent form 
\begin{align}
 \mathscr{A}_\mu 
 =& (\bm{n} \cdot \mathscr{A}_\mu)\bm{n} + \mathscr{A}_\mu - (\bm{n} \cdot \mathscr{A}_\mu) \bm{n}
 \nonumber\\
 =& (\bm{n} \cdot \mathscr{A}_\mu)\bm{n} + (\bm{n}\cdot\bm{n}) \mathscr{A}_\mu - (\bm{n} \cdot \mathscr{A}_\mu) \bm{n}
 \nonumber\\
 =& (\bm{n} \cdot \mathscr{A}_\mu)\bm{n} + \bm{n} \times (\mathscr{A}_\mu \times \bm{n})
 \nonumber\\
 =& (\bm{n} \cdot \mathscr{A}_\mu)\bm{n} - g^{-1}\bm{n} \times \partial_\mu \bm{n} + g^{-1}\bm{n} \times (\partial_\mu \bm{n} + \mathscr{A}_\mu \times \bm{n}) 
 \nonumber\\
 =& (\bm{n} \cdot \mathscr{A}_\mu)\bm{n} + g^{-1} \partial_\mu \bm{n} \times \bm{n} + g^{-1}\bm{n} \times D_\mu[\mathscr{A}]\bm{n}  ,
\label{CFNform}
\end{align}
where we have used only the relation $\bm{n}\cdot\bm{n}=1$ in the second equality. 

An important observation from (\ref{def:X}) is that {\it the local gauge transformations $\delta c_\mu$ and $\delta\mathbb X_\mu$ are uniquely determined  once  the transformations 
$\delta{\bm n}$ and $\delta\mathscr{A}_\mu$ are specified}.%
\footnote{
We emphasize this fact, which is pointed out in Ref.\cite{Shabanov99b}, because it is important from our viewpoint.
} 

\begin{figure}[htbp]
\begin{center}
\includegraphics[height=6.5cm]{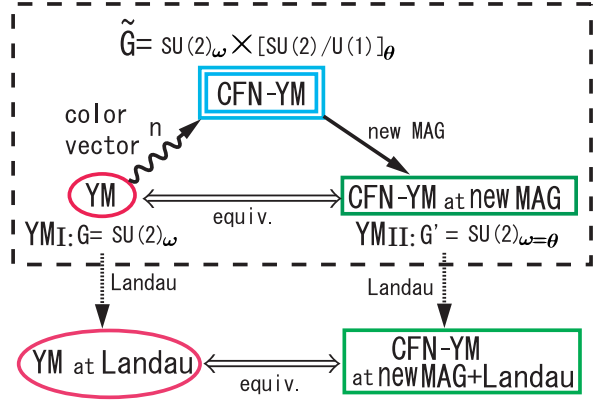}
\caption{\small 
The relationship between the CFN-Yang--Mills theory and the original Yang-Mills theory. The CFN-Yang--Mills theory obtained through the CFN decomposition has a  larger (local and global) gauge symmetry than the original Yang-Mills theory and becomes equivalent to the original Yang-Mills theory after the new MAG is imposed.
}
\label{fig:sym-cfn-ym}
\end{center}
\end{figure}

Now we consider the local gauge symmetry possessed by the Yang-Mills theory written in terms of CFN variables, which we call {\it CFN--Yang-Mills theory}.  
\begin{itemize}
\item
The invariance of the Lagrangian is guaranteed by the usual gauge transformation:
\begin{equation}
\delta\mathscr{A}_\mu(x)
  =D_\mu[\mathscr{A}]{\bm\omega}(x) .
\label{gtA}
\end{equation}
This symmetry is the local $G=SU(2)$ gauge symmetry and denoted $SU(2)_{local}^{\omega}$.

\item
The gauge transformation of $\bm{n}$ is nothing but the map from $S^2$ to $S^2$  at each spacetime point, because $\bm{n}$ is always defined to be a three-dimensional unit vector field, i.e., ${\bm n}(x)^2=1$. 
Therefore, it is  expressed as a local rotation by an angle ${\bm\theta}(x)$:%
\footnote{
Shabanov\cite{Shabanov99b} argued that it is possible to consider a more general transformation of the field ${\bm n}(x)$, even a nonlocal one, keeping the condition ${\bm n}(x)^2=1$.  However, it is unrealistic to consider an explicit transformation other than the local rotation treated in this paper. 
}
\begin{equation}
\delta{\bm n}(x)
  =g{\bm n}(x) \times {\bm\theta}(x) 
  =g{\bm n}(x) \times {\bm\theta}_\perp(x) ,
\label{gtn}
\end{equation}
where  
${\bm\theta}_\perp(x)$ is  perpendicular to $\bm{n}(x)$  [i.e., $\bm n(x)\cdot{\bm\theta}_\perp(x)=0$] and has two independent components.
For the parallel component, ${\bm\theta}_\parallel(x)=\theta_\parallel(x)\bm n(x)$, the vector field $\bm{n}(x)$ is invariant under this transformation [a  rotation about the axis of $\bm{n}(x)$]. Therefore, it is a {\it redundant} symmetry, say the U(1)$^{\theta}$ symmetry, of the CFN-Yang-Mills theory, as $c_\mu(x)$ and $\mathbb{X}_\mu(x)$ are also unchanged for a given $\mathscr{A}_\mu(x)$. (Note that $S^2 \simeq SU(2)/U(1)$.) 
Therefore, this symmetry is the local SU(2)/U(1) symmetry and denoted  $[SU(2)/U(1)]_{local}^{\theta}$.

\end{itemize}
Note that 
${\bm\omega}(x)$ and ${\bm\theta}(x)$ are independent, since 
the original Yang-Mills Lagrangian does not depend on the choice of ${\bm\theta}(x)$.
For later convenience, we distinguish between the above transformations by $\delta_\theta$ and $\delta_\omega$ as follows:
\begin{align}
 {\bm\theta}={\bm0} ,  {\bm\omega}\ne{\bm0}:
 \Longrightarrow & 
\delta_\omega{\bm n}(x)={\bm0},
\quad
\delta_\omega\mathscr{A}_\mu(x)
  =D_\mu[\mathscr{A}]{\bm\omega}(x) ,
\\
{\bm\theta}\ne{\bm0} (\bm{n}\cdot{\bm\theta}=0), {\bm\omega}={\bm0}:
 \Longrightarrow & 
\delta_\theta{\bm n}(x)
 =g{\bm n}(x)\times{\bm\theta}(x),
\quad
\delta_\theta\mathscr{A}_\mu(x)={\bm0}.
\end{align}
The general local gauge transformation in  the CFN--Yang-Mills theory is obtained by combining 
 $\delta_\theta$ and $\delta_\omega$. 
 Thus, {\it the CFN--Yang-Mills theory has the local gauge symmetry 
\begin{equation}
\tilde{G}^{\omega,\theta}_{local} := SU(2)_{local}^{\omega} \times [SU(2)/U(1)]_{local}^{\theta}  ,
\end{equation}
i.e., the direct product of $SU(2)_{local}^{\omega}$ and $[SU(2)/U(1)]_{local}^{\theta} $, which is larger than the local $G=SU(2)_{local}^{\omega}$ symmetry of the original Yang-Mills theory} (see Fig.~\ref{fig:sym-cfn-ym}).


In the papers \cite{BCK01,Cho03,Kondo04}, two local gauge transformations are introduced 
by decomposing the original gauge transformation, 
$
  \delta_\omega \mathscr{A}_\mu(x)  =  D_\mu[\mathscr{A}] \bm{\omega}(x) .
$

\underline{Local gauge transformation I}:  
\begin{subequations}
\begin{align}
  \delta_\omega \bm{n}  =& 0  ,
\\
 \delta_\omega c_\mu =& \bm{n} \cdot  D_\mu[\mathscr{A}] \bm{\omega} ,
\\
  \delta_\omega \mathbb{X}_\mu =&     D_\mu[\mathscr{A}] \bm{\omega} - \bm{n}( \bm{n} \cdot  D_\mu[\mathscr{A}] \bm{\omega}) ,
\end{align}
\end{subequations}

\underline{Local gauge transformation II}:  
\begin{subequations}
\begin{align}
  \delta_{\omega}' \bm{n}  =& g \bm{n} \times \bm{\omega'}  ,
\\
 \delta_{\omega}' c_\mu =&    \bm{n} \cdot \partial_\mu \bm{\omega'}   ,
\\
  \delta_{\omega}' \mathbb{X}_\mu =&  g \mathbb{X}_\mu \times \bm{\omega'} ,
\end{align}
\end{subequations}
The gauge transformation I has been called the passive or quantum gauge transformation, while II has been called the active or background gauge transformation.  However, this classification is not necessarily independent, and it leads to sometimes confusing and misleading results.
The local gauge transformation I defined in the previous paper \cite{Kondo04} is identical to $\delta_\omega$.
In order to see how the gauge transformation II defined in  \cite{Kondo04} is reproduced,  
we apply the gauge transformations (\ref{gtA}) and (\ref{gtn}) to   (\ref{def:X}).  Then we can show that the gauge transformation of other CFN variables are given by 
\footnote{
This transformation law was obtained by Shabanov \cite{Shabanov99b}. In it, $\delta \bm{n}(x)$ is not specified and is left undetermined on the right-hand side, based on the viewpoint that $\delta \bm{n}(x)$ should be determined by the choice of the constraint condition $\bm{\chi}(\bm{n},\mathscr{A}) \equiv \bm{\chi}(\bm{n},c,\mathbb{X})=0$, which reduces the degrees  of freedom to the original ones (by solving $\delta \bm{\chi}=0$ on the hypersurface $\bm{\chi}=0$). 
The new MAG defined below is consistent with the local rotation of $\bm{n}$, as a part of the gauge transformation II.  Therefore, our result is in agreement with the claim given in \cite{Shabanov99b} [see (\ref{dMAG}) and (\ref{gtF})]. 
}
\begin{align}
  \delta c_\mu(x) 
  =  g(\bm{n}(x) \times \mathscr{A}_\mu(x)) \cdot (\bm{\omega}_\perp(x) - \bm{\theta}_\perp(x)) + \bm{n}(x) \cdot \partial_\mu \bm{\omega}(x) ,
  \label{gtc}
\\
  \delta \mathbb{X}_\mu(x) 
  = g \mathbb{X}_\mu(x) \times  (\bm{\omega}_\parallel(x)+\bm{\theta}_\perp(x)) + D_\mu[\mathbb{V}](\bm{\omega}_\perp(x)-\bm{\theta}_\perp(x))  .
  \label{gtX}
\end{align}
If $\bm\omega_\perp(x)=\bm\theta_\perp(x)$,
the transformations (\ref{gtc}) and (\ref{gtX}) reduce to the gauge transformation II with the parameter $\bm\omega'(x)=(\bm\omega_\parallel(x),\bm\omega_\perp(x)=\bm\theta_\perp(x))$.
Therefore,  the gauge transformation II corresponds to the special case 
$\bm\omega_\perp(x)=\bm\theta_\perp(x)$.


\subsection{A new viewpoint for the CFN-Yang--Mills theory}


  The CFN-Yang-Mills theory  has the local gauge symmetry $\tilde{G}^{\omega,\theta}_{local}$ which is larger than that of the original Yang-Mills theory,
  because we can rotate the CFN variable $\bm{n}(x)$ by an angle $\bm{\theta}^{\perp}(x)$ independently of the gauge transformation parameter $\bm{\omega}(x)$ of  $\mathscr{A}_\mu(x)$.  
     In order to obtain the gauge theory with the same local gauge symmetry as the original Yang-Mills theory, therefore,  we proceed to impose a gauge fixing condition by which $\tilde{G}^{\omega,\theta}_{local}$ is broken down to $SU(2)$, a subgroup of $\tilde{G}^{\omega,\theta}_{local}$ (see Fig.~\ref{fig:sym-cfn-ym}).

We have found that one way of imposing such a gauge fixing condition is to impose the minimizing condition 
\begin{align}
 0 = \delta\int d^4x\frac12\mathbb X_\mu^2 ,
 \label{MAGcond}
\end{align}
{\it with respect to the enlarged gauge transformation} $(\omega, \theta)$, 
which we call the new maximal Abelian gauge (nMAG).
This is done as follows. 
Because  the relationship (\ref{XnA}) leads to 
\begin{align}
   g^2 \mathbb X_\mu^2
 =  ({\bm n}\times D_\mu[\mathscr{A}]{\bm n})^2
=      \left\{
     (D_\mu[\mathscr{A}]{\bm n})^2
     -({\bm n}\cdot D_\mu[\mathscr{A}]{\bm n})^2
    \right\}
=    (D_\mu[\mathscr{A}]{\bm n})^2 ,
\label{X2}
\end{align}
the local gauge transformation of $\mathbb X^2$ is calculated as
\begin{align}
\delta\frac12\mathbb X_\mu^2
  &=g^{-2}
     (D_\mu[\mathscr{A}]{\bm n}) \cdot \delta(D_\mu[\mathscr{A}]{\bm n})
    \nonumber \\
  &=g^{-2}
     (D_\mu[\mathscr{A}]{\bm n}) \cdot 
     (D_\mu[\mathscr{A}]\delta{\bm n}
      +g\delta\mathscr{A}_\mu\times\bm n)
    \nonumber \\
  &=g^{-2}
     (D_\mu[\mathscr{A}]{\bm n}) \cdot 
     \{gD_\mu[\mathscr{A}](\bm n\times\bm\theta_\perp)
       +(gD_\mu[\mathscr{A}]\bm\omega)\times\bm n\}
    \nonumber \\
  &=g^{-2}
     (D_\mu[\mathscr{A}]{\bm n}) \cdot 
     \{(gD_\mu[\mathscr{A}] \bm n) \times\bm\theta_\perp
      +{\bm n}\times (gD_\mu[\mathscr{A}]\bm\theta_\perp)
       -{\bm n} \times  (gD_\mu[\mathscr{A}]\bm\omega) \}
    \nonumber \\
  &=g^{-1}
     (D_\mu[\mathscr{A}]{\bm n}) \cdot 
     \{D_\mu[\mathscr{A}](\bm\omega-\bm\theta_\perp) \times \bm n \}
    \nonumber \\
  &= g^{-1}
     (D_\mu[\mathscr{A}]{\bm n}) \cdot 
     \{D_\mu[\mathscr{A}](\bm\omega_\perp - \bm\theta_\perp) \times \bm n \} ,
\label{eq:dX^2}
\end{align}
where we have used (\ref{gtn}) and (\ref{gtA}) in the third equality, 
and in the last equality   
we have decomposed $\bm\omega-\bm\theta_\perp$ into the parallel component $\bm\omega_\parallel= \omega_\parallel \bm n$ and perpendicular component $\bm\omega_\perp-\bm\theta_\perp$ and used the fact that the parallel part does not contribute, because
$
 D_\mu[\mathscr{A}] \bm\omega_\parallel  \times \bm n
=   D_\mu[\mathscr{A}]( \omega_\parallel \bm n)  \times \bm n
=   \{ \bm n \partial_\mu  \omega_\parallel  +  \omega_\parallel  D_\mu[\mathscr{A}]\bm n \}  \times \bm n 
=   \omega_\parallel   (D_\mu[\mathscr{A}]\bm n )  \times \bm n .
$
Therefore, the local gauge transformation II  does not change $\mathbb X^2$.

Then the average over the spacetime of  (\ref{eq:dX^2}) reads  
\begin{align}
\delta\int d^4x\frac12\mathbb X_\mu^2
  &=g^{-1} \int d^4x
     (D_\mu[\mathscr{A}]{\bm n}) \cdot 
     \{ D_\mu[\mathscr{A}](\bm\omega_\perp-\bm\theta_\perp) \times \bm n\}
    \nonumber \\
  &= \int d^4x
     \mathbb X_\mu\cdot
     D_\mu[\mathscr{A}](\bm\omega_\perp-\bm\theta_\perp)
    \nonumber \\
  &=- \int d^4x
     (\bm\omega_\perp-\bm\theta_\perp)\cdot
     D_\mu[\mathscr{A}]\mathbb X_\mu
    \nonumber \\
  &=- \int d^4x
     (\bm\omega_\perp-\bm\theta_\perp)\cdot
     D_\mu[\mathbb V]\mathbb X_\mu ,
 \label{minX2}
\end{align}
where we have used (\ref{def:X}) in the second equality and integration by parts in the third equality. 
Hence, the minimizing condition (\ref{MAGcond}) 
for arbitrary $\bm\omega_\perp$ and  $\bm\theta_\perp$ 
yields a gauge-fixing condition in differential form:%
\footnote{
Of course, it is trivial that two constraints are necessary and sufficient to eliminate the two extra degrees introduced by the CFN decomposition. 
Indeed, the form of the new MAG condition is the same as that   given in Ref.\cite{Shabanov99b}. 
However, there is no argument to identify which part of the enlarged gauge symmetry is fixed by this constraint, even in the case of a local rotation for $\bm{n}$. This is important to avoid the misunderstanding that appears in the literature. It is not the naively expected $[SU(2)/U(1)]^{\theta}_{local}$ symmetry that is fixed by two  constraint conditions, as is clear from our argument. The correct identification of the gauge symmetry influences the explicit form of the BRST transformation \cite{KMS05b}, and finding the correct way of implementing the CFN decomposition on a lattice \cite{KKMSS05,KKMSSI05}.
}
\begin{equation}
\mathbb F_{\rm MA} = \bm{\chi}
 :=D_\mu[\mathbb V]\mathbb X_\mu
 \label{dMAG}
 \equiv0 .
\end{equation}
Note that (\ref{dMAG}) denotes two conditions, since 
$\bm{n} \cdot \bm{\chi} =0$, which follows from the identity 
$ \bm{n} \cdot  D_\mu[\mathbb V]\mathbb X_\mu = 0$. 
Therefore, the minimization condition (\ref{MAGcond}) works as a gauge fixing condition, 
except in the case of the gauge transformation II, i.e., $\bm\omega_\perp(x)=\bm\theta_\perp(x)$.
In fact, the condition (\ref{dMAG}) does not transform covariantly, except in the case of the gauge transformation II, because 
 the gauge transformation of the condition (\ref{dMAG}) reads
\begin{align}
  \delta \bm{\chi}
  = g \bm{\chi} \times  (\bm{\omega}_\parallel +\bm{\theta}_\perp ) 
  - g^2 \mathbb{X}_\mu \times [\mathbb{X}_\mu \times (\bm{\omega}_\perp -\bm{\theta}_\perp )]
  + D_\mu[\mathbb{V}]D_\mu[\mathbb{V}](\bm{\omega}_\perp -\bm{\theta}_\perp )  .
  \label{gtF}
\end{align}
For  
$\bm\omega_\perp(x)=\bm\theta_\perp(x)$, the condition (\ref{dMAG}) transforms covariantly, because 
$
\delta \bm{\chi}
  = g \bm{\chi} \times  (\bm{\omega}_\parallel +\bm{\omega}_\perp ) 
  = g \bm{\chi} \times  \bm{\omega}.   
$
Here the local rotation of $\bm{n}$,  i.e., 
$
\delta{\bm n}(x)  =g{\bm n}(x) \times {\bm\theta}_\perp(x)
$,
leads to  $\delta \bm{\chi}=0$ on $\bm{\chi}=0$. 
Moreover, the U(1)$_{local}^{\omega}$ part in $G=SU(2)_{local}^{\omega}$ is not affected by this condition. Hence, the gauge symmetry corresponding to $\bm\omega_\parallel(x)$ remains unbroken. 

Therefore, if we impose the condition (\ref{MAGcond}) on the CFN-Yang--Mills theory,
we have a gauge theory with the local gauge symmetry 
$G'=SU(2)_{local}^{\omega=\theta}$ corresponding to the gauge transformation parameter $\bm\omega(x)=(\bm\omega_\parallel(x),\bm\omega_\perp(x)=\bm\theta_\perp(x))$, which is the diagonal SU(2) part $\tilde{G}^{\omega=\theta}_{local}$ of the original 
$\tilde{G}^{\omega,\theta}_{local}$.  
The local gauge symmetry $G'=SU(2)_{local}^{\omega=\theta}$ is the same as the gauge symmetry II.

The form of the condition (\ref{dMAG}) is identical to that of the MAG fixing condition for the CFN variables (see, e.g., \cite{Kondo04}). 
However, (\ref{dMAG}) and  (\ref{MAGcond}) are completely different from the conventional MAG fixing condition \cite{tHooft81,KLSW87}, which has been used  to this time to fix the off-diagonal part of the local gauge symmetry SU(2) of the original Yang-Mills theory (based on the Cartan decomposition) keeping the U(1) part intact, because the MAG introduced in this paper plays the role of eliminating the extra gauge symmetry generated by using the CFN variables and leaves the full SU(2) local gauge symmetry.  
Therefore, we call (\ref{MAGcond}) [and (\ref{dMAG})] the {\it new MAG} (the differential form).
\footnote{
Note that (\ref{MAGcond}) is more general than (\ref{dMAG}), since (\ref{dMAG}) is the differential form, which is valid only in the absence of Gribov copies.  
The condition (\ref{MAGcond}) is the most general MAG condition, which can be used also in numerical simulations on a lattice and works even if Gribov copies exist and leads to the true minimum, while (\ref{dMAG}) leads only to the local minimum along the gauge orbit. 
}
The new Yang-Mills theory obtained by imposing the nMAG on the CFN-Yang-Mills theory is called  {\it Yang--Mills theory II}  hereafter. 
Among the three gauge degrees of freedom $\bm{\omega}=(\bm{\omega}_\perp,\bm{\omega}_\parallel)$ and two degrees of freedom $\bm{\theta}_\perp$ in the CFN-Yang--Mills theory, two extra gauge degrees of freedom  were eliminated by imposing the two conditions expressed by the nMAG, $\mathbb F_{\rm MA}=0$, and then the remaining degrees of freedom in  Yang--Mills theory II are those in the same as the original Yang-Mills theory I. 
In the Yang--Mills theory II, we can impose any further gauge-fixing condition for fixing the diagonal SU(2) after the nMAG is imposed, e.g., instead of Landau in Fig.~\ref{fig:sym-cfn-ym}. In fact, we can furthermore impose the conventional MAG, if desired.  
 This leads to the possibility of examining the gauge invariance even after the nMAG. In the previous approach, the MAG is one of the gauge fixings and there is no specific reason to take the MAG (except for the coincidence of the degrees of freedom).  But in our approach the MAG plays a different and distinguished role.  Even after imposing the nMAG, Yang--Mills theory II has the full SU(2) symmetry.

Our viewpoint for the CFN-Yang--Mills theory resolves in a natural way the crucial issue of the discrepancy in the independent degrees of freedom between the two theories, i.e., the original Yang-Mills theory and the CFN-Yang--Mills theory.  
Moreover, it reveals the necessity of adopting the nMAG in the CFN-Yang--Mills theory, although the conventional MAG is merely one choice of the gauge fixings. 
To the best of our knowledge to this time, this point had not been correctly understood.

In the paper \cite{KKMSS05,KKMSSI05}, we performed Monte Carlo simulations of the CFN-Yang-Mills theory for the first time, by imposing the nMAG and the Lattice--Landau gauge (LLG) simultaneously.%
\footnote{
A possible algorithm for the numerical simulation was proposed in \cite{Shabanov01}, and actual simulations were first attempted in \cite{DHW02}.  However, from our point of view, they cannot be identified with the CFN-Yang-Mills theory.   
Here we point out that only the  field $\bm{n}$ was constructed in these works, and the simulation results show the breaking of the global SU(2) invariance, even in the Landau gauge, which cannot be regarded as the correct implementation of the CFN decomposition on a lattice. It is the essence to preserve the color symmetry. 
Details are given in Refs.\cite{KKMSS05,KKMSSI05}. 
}
Here, the LLG fixes the local gauge symmetry $G'=SU(2)_{local}^{\omega=\theta}$ and the MAG imposed in Refs.\cite{KKMSS05,KKMSSI05} is the nMAG mentioned above, not the conventional MAG. 
In general, we can impose any gauge fixing condition instead of LLG, in addition to the nMAG in  numerical simulations. 
This is an advantage of our viewpoint for the CFN-Yang--Mills theory. 

 Yang--Mills theory II was constructed on a vacuum selected in a gauge invariant way among the possible vacua of the CFN-Yang--Mills theory, since the nMAG is satisfied for the CFN field configurations realizing the minimum of the functional $\int d^4x\frac12\mathbb X_\mu^2$, and the minimum 
\begin{align}
 \min_{\bm\omega,\bm\theta} \int d^4x\frac12\mathbb X_\mu^2 
\end{align}
 is gauge invariant in the sense that it does no longer change the value with respect to the enlarged local gauge transformation. 
Therefore, {\it the nMAG is a gauge-invariant criterion for choosing a vacuum on which  Yang--Mills theory II is defined from the vacua of the CFN-Yang--Mills theory}, although the nMAG is not necessarily a unique prescription for selecting  the gauge-invariant vacuum.  
This demonstrates the quite different role played by the nMAG compared with the conventional MAG. 
The original Yang-Mills theory with the local gauge symmetry $G=SU(2)_{local}^{\omega}$, i.e., Yang-Mills theory I, is reproduced from the CFN-Yang--Mills theory by fixing the field variable $\bm{n}(x)$ as  $\bm{n}(x) \equiv \bm{n}_{\infty}:=(0,0,1)$ at the all spacetime points. 


\subsection{Independent variables of the respective theory}


In order to clarify which variables are independent variables in the respective theory, we write the partition function of the respective theory with the integration measure, up to the gauge fixing term and the associated Faddeev--Popov ghost term to be investigated in \cite{KMS05b}.

 The partition function of the original Yang--Mills theory (Yang--Mills I) is   in the Euclidean formulation:
\begin{align}
 Z_{{\rm YM}} = \int \mathcal{D}\mathscr{A}_\mu \exp (-S_{{\rm YM}}[\mathscr{A}]) .
\end{align}
By introducing the auxiliary color field $\bm{n}(x)$, 
the CFN-Yang--Mills theory is first defined by a partition function written in terms of $\bm{n}(x)$ and $\mathscr{A}_\mu(x)$,
\begin{align}
 \tilde{Z}_{{\rm YM}}  = \int \mathcal{D}\bm{n}  \delta(\bm n\cdot\bm n-1)
\int \mathcal{D}\mathscr{A}_\mu \exp (-S_{{\rm YM}}[\mathscr{A}]) , 
\label{Z}
\end{align}
and then it is rewritten in terms of the CFN variables
$
(\bm n,c_\mu,\mathbb X_\mu)
$
as 
\begin{align}
 \tilde{Z}_{{\rm YM}}  
= \int \mathcal{D}\bm{n}  \delta(\bm n\cdot\bm n-1)
 \int \mathcal{D}c_\mu \int 
\mathcal{D}\mathbb{X}_\mu \delta(\bm n\cdot\mathbb X_\mu)  
J 
 \exp (-\tilde S_{\rm YM}[\bm n,c,\mathbb X]) , 
\end{align}
where $J$ is the Jacobian associated with the change of variables from 
$
(\bm n, \mathscr A_\mu)
$ 
to 
$
(\bm n,c_\mu,\mathbb X_\mu)
$
and the action $\tilde S_{\rm YM}[\bm n,c,\mathbb X]$ is obtained by substituting the CFN decomposition of $\mathscr A_\mu$
 into $S_{{\rm YM}}[\mathscr{A}]$: 
\begin{align}
\tilde S_{\rm YM}[\bm n,c,\mathbb X]
 =S_{\rm YM}[\mathscr A] .
\end{align}

In order to fix the enlarged symmetry in the CFN-Yang--Mills theory and retain only the gauge symmetry II,
we impose the constraint $\bm{\chi}[ \mathscr{A},\bm{n}]=0$ (the new maximal Abelian gauge).
Then, we write unity in the form 
\begin{align}
  1 = \int \mathcal{D} \bm{\chi}^\theta \delta(\bm{\chi}^\theta)
=   \int\! \mathcal{D}\bm\theta\delta(\bm\chi^\theta)
   \det\left(\frac{\delta\bm\chi^\theta}{\delta{\bm\theta}}\right) ,
\end{align}
where $\bm{\chi}^\theta$ is the constraint written in terms of the gauge-transformed variable, i.e., 
$\bm{\chi}^\theta:=\bm{\chi}[ \mathscr{A},\bm{n}^\theta]$,
and then we insert this into the functional integral   (\ref{Z}).
This yields
\begin{align}
 Z_{{\rm YM}} = \int \mathcal{D}\bm{n}  \delta(\bm n\cdot\bm n-1) \int \mathcal{D}\mathscr{A}_\mu \int\! \mathcal{D}\bm\theta\delta(\bm\chi^\theta)
   \det\left(\frac{\delta\bm\chi^\theta}{\delta{\bm\theta}}\right) 
\exp (-S_{{\rm YM}}[\mathscr{A}]) .
\end{align}
Then  we cast the partition function of the CFN-Yang--Mills theory II   into the form%
\footnote{
It is not difficult to show that the Jacobian $J$ for the change of variables from ${\bm n}^A$ and $\mathscr{A}_\mu^A$ to ${\bm n}^A, c_\mu$ and $\mathbb{X}_\mu^A$ is equal to 1, if the integration measures $\mathcal{D}\bm{n}$ and $\mathcal{D}\mathbb{X}_\mu$ are understood to be written in terms of independet degrees of freedom by taking into account the constraints $\bm n\cdot\bm n=1$ and $\bm n\cdot\mathbb X_\mu=0$.  In fact, only the independent degrees of freedom have been used to calculate the explicit form of the effective potential \cite{Cho03,Kondo04}. 
Therefore, we do not pay special attention to the Jacobian $J$ in what follows.
}
\begin{align}
 \tilde{Z}_{{\rm YM}}  
=& \int \mathcal{D}\bm{n}  \delta(\bm n\cdot\bm n-1)
 \int \mathcal{D}c_\mu \int 
\mathcal{D}\mathbb{X}_\mu \delta(\bm n\cdot\mathbb X_\mu)  
J  
\nonumber\\
 & \times \int\! \mathcal{D}\bm\theta\delta(\bm\chi^\theta)
   \det\left(\frac{\delta\bm\chi^\theta}{\delta{\bm\theta}}\right)
 \exp (-\tilde S_{\rm YM}[\bm n,c,\mathbb X]) . 
\end{align}
We next perform the change of variables $\bm{n} \rightarrow \bm{n}^{\theta}$ obtained through a local rotation by the angle $\theta$ and the corresponding gauge   transformations II for the other CFN variables $c_\mu$ and $\mathbb X_\mu$: 
$c_\mu, \mathbb{X}_\mu \rightarrow c_\mu^{\theta}, \mathbb{X}_\mu^{\theta}$. 
From the gauge invariance II of the action $\tilde S_{\rm YM}[\bm n,c,\mathbb X]$ and the measure 
$
 \mathcal{D}\bm{n}  \delta(\bm n\cdot\bm n-1)
  \mathcal{D}c_\mu  
\mathcal{D}\mathbb{X}_\mu \delta(\bm n\cdot\mathbb X_\mu)
$, 
we can rename the dummy integration variables $\bm{n}^{\theta}, c_\mu^{\theta}, \mathbb{X}_\mu^{\theta}$  as $\bm{n}, c_\mu, \mathbb{X}_\mu$.
Thus the integrand does not depend on $\theta$, and the gauge volume $\int\! \mathcal{D}\bm\theta$ can be removed: 
\begin{align}
 \tilde{Z}_{{\rm YM}}  
=& \int\! \mathcal{D}\bm\theta 
\int \mathcal{D}\bm{n}  \delta(\bm n\cdot\bm n-1)
 \int \mathcal{D}c_\mu \int 
\mathcal{D}\mathbb{X}_\mu \delta(\bm n\cdot\mathbb X_\mu)  
J  
\nonumber\\
 & \times \delta(\bm\chi)
   \det\left(\frac{\delta\bm\chi}{\delta{\bm\theta}}\right)
 \exp (-\tilde S_{\rm YM}[\bm n,c,\mathbb X]) .
\end{align}
Note that the Faddeev--Popov determinant $\det\left(\frac{\delta\bm\chi}{\delta{\bm\theta}}\right)$ can be rewritten into another form,   
$
 \Delta_{FP}^{nMAG}
:= \det\left(\frac{\delta\bm\chi}{\delta{\bm\theta}}\right)_{\bm{\chi}=0}
=   \det\left(\frac{\delta\bm\chi}{\delta\bm n^\theta}\right)_{\bm{\chi}=0}.
$
This is the same as the determinant called the Shabanov determinant \cite{Shabanov99b}, 
$
\Delta_{S}[\mathscr{A}_\mu, \bm{n}]  
:= \det \left|{\delta \bm{\chi} \over \delta \bm{n}} \right|_{\bm{\chi}=0} ,
$
which gurantees the equivalence of Yang-Mills theories I and II. 
From our viewpoint, therefore, the Shabanov determinant is simply the Faddeev--Popov determinant associated with the nMAG.  
Thus, the the partition function of Yang-Mills theory II is  given by
\begin{align}
 Z_{{\rm YM}}^\prime  
=& \int \mathcal{D}\bm{n}  \delta(\bm n\cdot\bm n-1)
 \int \mathcal{D}c_\mu \int 
\mathcal{D}\mathbb{X}_\mu \delta(\bm n\cdot\mathbb X_\mu)  
J 
\nonumber\\ & \times  
\delta(\tilde{\bm\chi})  
   \Delta_{FP}^{nMAG}
 \exp (-\tilde S_{\rm YM}[\bm n,c,\mathbb X]) , 
\end{align}
where the constraint is written in terms of the CFN variables:
\begin{equation}
\tilde{\bm\chi} 
 :=\tilde{\bm\chi} [\bm n, c,\mathbb X]
 :=D^\mu[\mathbb V ]\mathbb X_\mu, \quad
\mathbb V_\mu \equiv c_\mu  \bm{n} 
  +g^{-1}\partial_\mu \bm{n} \times \bm{n} .
\end{equation}
In Yang--Mills theory II, the independent variables are regarded as $\bm{n}(x)$, $c_\mu(x)$ and $\mathbb{X}_\mu(x)$. 

In order to obtain a completely gauge-fixed theory, we must repeat the gauge-fixing procedures after imposing both the nMAG and the gauge fixing condition  for SU(2) symmetry, e.g., the Landau gauge $\partial^\mu \mathscr{A}_\mu(x)=0$.
According to the clarification of the symmetry in the CFN-Yang--Mills theory explained above,  we can obtain the unique Faddeev-Popov ghost terms associated with the gauge fixing conditions adopted in quantization.  
This is  another advantage of our viewpoint for the CFN-Yang--Mills theory. The explicit derivation of the FP ghost term has been worked out in a separate paper \cite{KMS05b}.


\subsection{Gauge invariance and observables}


The above consideration shows that 
the gauge invariant quantities in Yang--Mills theory II must be regarded as those which are invariant under the gauge transformation II. 
In this sense, $\mathbb{X}_\mu(x)^2$ is a gauge invariant quantity and must have a definite physical meaning. 
Therefore, the vacuum condensation 
$\left< \mathbb{X}_\mu(x)^2 \right>$ could be an important physical quantity. 
In fact, recalling that the Skyrme--Faddeev model \cite{FN97,MS04} is derived from this vacuum condensation as pointed out in \cite{Kondo04}, the vacuum condensation of mass dimension two, $\left< \mathbb{X}_\mu(x)^2 \right>$, could be related to this observable \cite{GSZ01,Kondo01,Slavnov04,Slavnov05,Kondo05}.

Surprisingly, we can write the gauge-invariant mass term using the CFN variables,
\begin{align}
  \mathscr{L}_{M}' := \frac{1}{2} M^2 \mathbb{X}_\mu(x)^2 , 
\end{align}
in Yang--Mills theory II  with the gauge symmetry II. 

In  the CFN-Yang--Mills theory with the enlarged gauge symmetry, 
this term can be identified with the kinetic term of the scalar field $\bm{\phi}(x)$ in the adjoint representation (\ref{X2}), 
\begin{align}
  \tilde{\mathscr{L}}_{M} = \frac{1}{2}  (D_\mu[\mathscr{A}] \bm{\phi}(x))^2 ,
  \quad \bm{\phi}(x) := g^{-1}M \bm{n}(x)  .
\end{align}
This can also be rewritten as a the gauge-invariant mass term similar to the St\"uckelberg type,
\begin{align}
  \tilde{\mathscr{L}}_{M} =& 
  \frac{1}{2}  M^2 (\mathscr{A}_\mu(x) - \mathbb{V}_\mu(x))^2 , 
\nonumber\\ 
  \mathbb{V}_\mu(x)  
    =& \bm{n}(x) [{\bm n}(x)\cdot\mathscr{A}_\mu(x)] 
  +g^{-1}\partial_\mu \bm{n}(x)\times \bm{n}(x) ,
\end{align}
where $\mathbb{V}_\mu(x)$ is written in terms of $\bm{n}(x)$ and the original Yang-Mills field $\mathscr{A}_\mu(x)$.  
The $\bm{n}(x)$ field is written as a quite complicated composite field of the original Yang-Mills field $\mathscr{A}_\mu(x)$, after the gauge fixing (see \cite{KKMSS05,KKMSSI05}).  
Further details will be discussed elsewhere. 


\subsection{Global gauge symmetry}


 We have imposed the nMAG to fix the off-diagonal symmetry of the local gauge symmetry $\tilde{G}^{\omega,\theta}_{local}$ and to keep the local gauge symmetry $SU(2)_{local}^{\omega=\theta}$.
Moreover, the nMAG also breaks the global gauge symmetry $\tilde{G}^{\omega,\theta}_{global}:=SU(2)_{global}^{\omega} \times [SU(2)/U(1)]_{global}^{\theta}$ into $SU(2)_{global}^{\omega=\theta}$.  In fact, the global parameters $\bm\theta$ and $\bm\omega$ yield  the change 
\begin{align}
\delta\frac12\mathbb X_\mu^2(x)
  &= 
     (D_\mu[\mathscr{A}]{\bm n}(x)) \cdot 
     \{ [ \mathscr{A}_\mu(x) \times (\bm\omega^\perp - \bm\theta^\perp)] \times  \bm n(x)  \}
\nonumber\\
  &= 
     (D_\mu[\mathscr{A}]{\bm n}(x)) \cdot (\bm\omega^\perp - \bm\theta^\perp)
      [{\bm n}(x) \cdot  \mathscr{A}_\mu(x)] 
 ,
\label{eq:dX^2b}
\end{align}
and the right-hand side is nonzero in general, since $D_\mu[\mathscr{A}]{\bm n}(x)$ is perpendicular to ${\bm n}(x)$.
Therefore, Yang-Mills theory II, i.e., the CFN-Yang-Mills theory with the nMAG, has the local gauge symmetry $SU(2)_{local}^{\omega=\theta}$ as well as the global gauge symmetry $SU(2)_{global}^{\omega=\theta}$.  These are the same local and global gauge symmetries as in the original Yang-Mills theory. 

Moreover, we can impose one more gauge fixing condition to fix the remaining local gauge symmetry $SU(2)_{local}^{\omega=\theta}$, so that it  maintains the global symmetry $SU(2)_{global}^{\omega=\theta}$, e.g., the Landau gauge. 
After imposing the nMAG and one more gauge fixing condition, the original local gauge symmetry, $\tilde{G}^{\omega,\theta}_{local}$, is completely fixed, while the global symmetry, $SU(2)_{global}^{\omega=\theta}$, is left intact. 



We must focus on the quantities which are invariant under $SU(2)_{global}^{\omega=\theta}$ from the viewpoint of color confinement.    In other words, only the $SU(2)_{global}^{\omega=\theta}$ singlet in CFN-Yang--Mills theory can have  physical meaning.  In fact, we have measured only the $SU(2)_{global}$ invariant quantities in the numerical simulations based on a new algorithm preserving  $SU(2)_{global}^{\omega=\theta}$ in Refs.\cite{KKMSS05,KKMSSI05}. 
 In general, the spontaneous breakdown of the color symmetry  $SU(2)_{global}^{\omega=\theta}$ could occur. 
In this case, we do not know what happens in the CFN-Yang--Mills theory and how the equivalence between the two theories is modified.  This issue should be investigated in subsequent works.


\section{Conclusion and discussion}
\setcounter{equation}{0}


We have shown that the CFN-Yang-Mills theory (the Yang-Mills theory written in terms of the CFN variables $\bm{n},c_\mu$ and $\mathbb{X}_\mu$) has the local gauge symmetry 
$\tilde{G}=SU(2)_{local}^{\omega} \times [SU(2)/U(1)]_{local}^{\theta}$, which is larger than the local gauge symmetry $G=SU(2)_{local}$ of the original Yang-Mills theory. 
We have imposed the new MAG to reduce the local gauge symmetry to the diagonal part, i.e.,  
$G'=SU(2)_{local}^{\omega=\theta}$.   
This procedure explicitly breaks the global gauge symmetry $SU(2)_{global}^{\omega} \times [SU(2)/U(1)]_{global}^{\theta}$ to
$SU(2)_{global}^{\omega=\theta}$ simultaneously.  
Then  Yang-Mills theory II, i.e., the CFN-Yang-Mills theory at the new MAG, has the same local and global gauge symmetries as the original Yang-Mills theory, before the conventional gauge fixing is imposed. 
The new MAG is used as a criterion for choosing the vacuum of Yang-Mills theory II in a gauge invariant way from the vacua of the CFN-Yang--Mills theory. 

The local gauge symmetry $G'=SU(2)^{\omega=\theta}$ of  Yang-Mills theory II is identical to the gauge symmetry II defined in \cite{Kondo04}, i.e., a local rotation of $\bm{n}$, $\mathbb{X}_\mu$ and an Abelian-like gauge transformation of $c_\mu$.  Therefore, the transformation properties of the CFN variables under the gauge symmetry II can lead to a new set of gauge invariant operators and observables (vacuum condensates) which were previously unexpected, e.g., a gauge-invariant mass term 
$
   \frac{1}{2} M^2 \mathbb{X}_\mu(x)^2 
$
and a composite operator of mass dimension two, $\mathbb{X}_\mu(x)^2$, and its condensate, $\left< \mathbb{X}_\mu(x)^2 \right>$. 

To quantize Yang-Mills theory II, we must introduce an appropriate gauge fixing condition to fix the gauge symmetry II, 
$G'=SU(2)_{local}^{\omega=\theta}$,  in the conventional sense.
We can definitely obtain the gauge-fixing term and the associated Faddeev--Popov ghost term.  
For example, we can choose the Landau gauge as the gauge-fixing condition that keeps the global gauge symmetry $SU(2)_{global}^{\omega=\theta}$ unbroken.  
In fact, we have tested this framework by Monte Carlo simulations on a lattice \cite{KKMSS05,KKMSSI05} and succeeded in extracting  novel nonperturbative features of the Yang-Mills theory. 
The conventional MAG could have been used at this stage breaking the global gauge symmetry.  Therefore, the new MAG introduced above is completely different from the conventional MAG. 
The new MAG is a logical necessity, while the conventional MAG is merely one of a number of possible gauge-fixing choices. 
This may shed new light on the role of MAG in Yang-Mills theory.

\section*{Acknowledgments}
This work is supported by a
Grant-in-Aid for Scientific Research (C)14540243 from the Japan Society for the Promotion of Science (JSPS), 
and in part by a Grant-in-Aid for Scientific Research on Priority Areas (B)13135203 from
the Ministry of Education, Culture, Sports, Science and Technology (MEXT).

\appendix

\section{Local Gauge Transformations I and II}\label{Appendix:gt}


For the CFN decomposition of the gauge field (\ref{CFN}), 
the non-Abelian field strength $\mathscr{F}_{\mu\nu}(x)$ is decomposed as
\begin{align}
  \mathscr{F}_{\mu\nu} [\mathscr{A}]
:=& \partial_\mu \mathscr{A}_\nu - \partial_\nu \mathscr{A}_\mu + g \mathscr{A}_\mu \times \mathscr{A}_\nu
\nonumber\\
=& \mathscr{F}_{\mu\nu} [\mathbb{V}] + \mathscr{F}_{\mu\nu} [\mathbb{X}] + g \mathbb{V}_\mu \times \mathbb{X}_\nu + g \mathbb{X}_\mu \times \mathbb{V}_\nu 
\nonumber\\
=& \mathscr{F}_{\mu\nu} [\mathbb{V}] +   D_\mu[\mathbb{V}] \mathbb{X}_\nu - D_\nu[\mathbb{V}] \mathbb{X}_\mu + g \mathbb{X}_\mu \times \mathbb{X}_\nu ,
\end{align}
where 
$
  D_\mu[\mathbb{V}]  := \partial_\mu   + g \mathbb{V}_\mu \times   
$
is the covariant derivative in the background field $\mathbb{V}_\mu$.

The field strength $\mathscr{F}_{\mu\nu} [\mathbb{V}]$  is further decomposed as
\begin{align}
\mathscr{F}_{\mu\nu} [\mathbb{V}]   
=& \mathscr{F}_{\mu\nu} [\mathbb{B}] + \mathscr{F}_{\mu\nu} [\mathbb{C}] + g \mathbb{B}_\mu \times \mathbb{C}_\nu + g \mathbb{C}_\mu \times \mathbb{B}_\nu 
\nonumber\\
:=& \mathbb{H}_{\mu\nu} + \mathbb{E}_{\mu\nu} :=  \mathbb{G}_{\mu\nu} , 
\end{align}
where the two kinds of field strength are defined by
\begin{align}
  \mathbb{H}_{\mu\nu}  :=& \mathscr{F}_{\mu\nu} [\mathbb{B}]
= \partial_\mu \mathbb{B}_\nu - \partial_\nu \mathbb{B}_\mu + g \mathbb{B}_\mu \times \mathbb{B}_\nu
\\
  \mathbb{E}_{\mu\nu} :=& \mathscr{F}_{\mu\nu} [\mathbb{C}] + g \mathbb{B}_\mu \times \mathbb{C}_\nu + g \mathbb{C}_\mu \times \mathbb{B}_\nu 
 .
\end{align}
Due to the special definition of $\mathbb{B}_\mu$,  the 'magnetic field' strength  $\mathbb{H}_{\mu\nu}$ is rewritten as
\begin{align}
  \mathbb{H}_{\mu\nu}  
&=   - g \mathbb{B}_\mu \times \mathbb{B}_\nu 
=  - g^{-1}   (\partial_\mu \bm{n} \times \partial_\nu \bm{n})  
=  H_{\mu\nu} \bm{n}, 
\\
H_{\mu\nu} &:=  - g^{-1} \bm{n} \cdot (\partial_\mu \bm{n} \times \partial_\nu \bm{n})  ,
\end{align}
where we have used the fact that $\mathbb{H}_{\mu\nu}$ is parallel to $\bm{n}$. 
Similarly, the 'electric field' strength $\mathbb{E}_{\mu\nu}$ is parallel to $\bm{n}$: 
\begin{align}
  \mathbb{E}_{\mu\nu} =  E_{\mu\nu} \bm{n}, \quad
E_{\mu\nu} := \partial_\mu c_\nu - \partial_\nu c_\mu .
\end{align}

The gauge transformations of the CFN variables are given as follows. 

\underline{Local gauge transformation I}   (the passive or quantum gauge transformation):
\begin{subequations}
\begin{align}
  \delta_\omega \bm{n}  =& 0  ,
\\
 \delta_\omega c_\mu =& \bm{n} \cdot  D_\mu[\mathscr{A}] \bm{\omega} ,
\\
  \delta_\omega \mathbb{X}_\mu =&     D_\mu[\mathscr{A}] \bm{\omega} - \bm{n}( \bm{n} \cdot  D_\mu[\mathscr{A}] \bm{\omega}) ,
\\ 
  \Longrightarrow  & \delta_\omega \mathbb{B}_\mu  =   0 ,
\quad
  \delta_\omega \mathbb{V}_\mu 
=   \bm{n}( \bm{n} \cdot  D_\mu[\mathscr{A}] \bm{\omega})  . 
\end{align}
\end{subequations}

The gauge transformation for the field strength can be obtained in the similar way: 
\begin{align}
  \delta_\omega \mathbb{E}_{\mu\nu} 
=& \bm{n} \delta_\omega {E}_{\mu\nu} 
= \bm{n} \{ \partial_\mu (\bm{n} \cdot D_\nu[\mathscr{A}] \bm{\omega}) - \partial_\nu (\bm{n} \cdot D_\mu[\mathscr{A}] \bm{\omega}) \} , 
\\
  \delta_\omega \mathbb{H}_{\mu\nu} 
=&  \bm{n} \delta_\omega {H}_{\mu\nu}  = 0 . 
\end{align}

\underline{Local gauge transformation II} (the active or background gauge transformation):
\begin{subequations}
\begin{align}
  \delta_{\omega}' \bm{n}  =& g \bm{n} \times \bm{\omega'}  ,
\\
 \delta_{\omega}' c_\mu =&    \bm{n} \cdot \partial_\mu \bm{\omega'}   ,
\\
  \delta_{\omega}' \mathbb{X}_\mu =&  g \mathbb{X}_\mu \times \bm{\omega'} ,
\\
\Longrightarrow  & \delta_{\omega}' \mathbb{B}_\mu  
=   D_\mu[\mathbb{B}] \bm{\omega'} - (\bm{n} \cdot \partial_\mu \bm{\omega'}) \bm{n}  ,
\quad
\delta_{\omega}' \mathbb{V}_\mu =   D_\mu[\mathbb{V}] \bm{\omega'}    . 
\end{align}
\end{subequations}
The gauge transformation for the field strength can be obtained in a similar way. 
It is easy to show that $\mathbb{G}_{\mu\nu}:=\mathscr{F}_{\mu\nu} [\mathbb{V}]$ (the sum $\mathbb{E}_{\mu\nu}+\mathbb{H}_{\mu\nu}$) is subject to the adjoint rotation 
\begin{align}
  \delta_\omega' \mathbb{G}_{\mu\nu} 
=& g \mathbb{G}_{\mu\nu} \times \bm{\omega}' ,
\end{align}
while this is not the case for the individual quantities, $E_{\mu\nu}$ and $H_{\mu\nu}$:
\begin{align}
  \delta_\omega' \mathbb{E}_{\mu\nu} 
=& g \mathbb{E}_{\mu\nu} \times \bm{\omega}' + \bm{n} \{ \partial_\mu (\bm{n} \cdot \partial_\nu \bm{\omega}') - \partial_\nu (\bm{n} \cdot \partial_\mu \bm{\omega}') \} , 
\\
  \delta_\omega' \mathbb{H}_{\mu\nu} =& g \mathbb{H}_{\mu\nu} \times \bm{\omega}' - \bm{n} \{ \partial_\mu (\bm{n} \cdot \partial_\nu \bm{\omega}') - \partial_\nu (\bm{n} \cdot \partial_\mu \bm{\omega}') \} . 
\end{align}
Hence, the squared field strength has the   SU(2)$_{\rm II}$ invariance 
\begin{align}
 \delta_\omega' \mathbb{G}_{\mu\nu}^2 
=&  0 .
\end{align}
The inner product of $\mathbb{G}_{\mu\nu}$ with $\bm{n}$ is also   SU(2)$_{\rm II}$ invariant:
\begin{align}
 \delta_\omega'( \bm{n} \cdot  \mathbb{G}_{\mu\nu}) 
 & \equiv \delta_\omega' G_{\mu\nu} 
=    0 ,
\\
 G_{\mu\nu} &=  \partial_\mu c_\nu - \partial_\nu c_\mu  -  g^{-1}  \bm{n} \cdot  (\partial_\mu \bm{n} \times \partial_\nu \bm{n}), \quad
c_\mu 
  ={\bm n} \cdot \mathscr{A}_\mu  .
\end{align}
This is not the case for the individual quantities, $E_{\mu\nu}$ and $H_{\mu\nu}$.

Moreover, we can show that 
\begin{align}
 \delta_\omega' ( D_\mu[\mathbb{V}] \mathbb{X}_\nu - D_\nu[\mathbb{V}] \mathbb{X}_\mu) 
=& g (D_\mu[\mathbb{V}] \mathbb{X}_\nu - D_\nu[\mathbb{V}] \mathbb{X}_\mu) \times \bm{\omega}' ,
\\
 \delta_\omega' (\mathbb{X}_\mu \times \mathbb{X}_\nu) =&   g (\mathbb{X}_\mu \times \mathbb{X}_\nu) \times \bm{\omega}' ,
\end{align}
Therefore, all the inner products among 
\begin{align}
\bm{n},  \quad
\mathbb{G}_{\mu\nu}:=\mathbb{E}_{\mu\nu}+\mathbb{H}_{\mu\nu}, \quad
\mathbb{X}_\mu \times \mathbb{X}_\nu
&\leftarrow \text{parallel~to~} \bm{n}
\\
\mathbb{X}_\mu,  \quad
( D_\mu[\mathbb{V}] \mathbb{X}_\nu - D_\nu[\mathbb{V}] \mathbb{X}_\mu),
&\leftarrow \text{perpendicular~to~} \bm{n}
\end{align}

\noindent
possess  SU(2)$_{\rm II}$ gauge invariance.
For example, we have
\begin{align}
 \delta_\omega' (D_\mu[\mathbb{V}] \mathbb{X}_\nu - D_\nu[\mathbb{V}] \mathbb{X}_\mu)^2 
=  0 ,
\quad 
 \delta_\omega' (\mathbb{X}_\mu \times \mathbb{X}_\nu)^2 =  0 ,
\\
  \delta_\omega'[\mathbb{G}_{\mu\nu} \cdot (g \mathbb{X}_\mu \times \mathbb{X}_\nu)] 
= 0. 
\end{align}

In particular, when $\bm{\omega}'(x)$ is parallel to $\bm{n}$, i.e., $\bm{\omega}'(x)=\theta'(x) \bm{n}(x)$, we obtain 
\noindent
\underline{Local U(1) gauge transformation II}:
\begin{subequations}
\begin{align}
  \delta_{\theta}' \bm{n}  =& 0  ,
\\
 \delta_{\theta}' c_\mu =&     \partial_\mu \theta'   ,
\\
  \delta_{\theta}' \mathbb{X}_\mu =&  g \mathbb{X}_\mu \times \theta'  \bm{n} ,
\\
\Longrightarrow  & \delta_{\theta}' \mathbb{B}_\mu  
=   0  ,
\quad
\delta_{\theta}' \mathbb{V}_\mu =    \bm{n} \partial_\mu \theta'    . 
\end{align}
\end{subequations}
Note that $\bm{n}$ and $\mathbb{B}_\mu$ are invariant under the U(1)$_{\rm II}$ gauge transformation, while $c_\mu$ transforms as the U(1)$_{\rm II}$ gauge field.  
It is easy to show the local  U(1)$_{\rm II}$ gauge invariance of the field strengths, i.e., 
\begin{align}
  \delta_{\theta}' \mathbb{E}_{\mu\nu}  =  0 , 
\quad
  \delta_{\theta}' \mathbb{H}_{\mu\nu} =  0 ,
\end{align}
which is also consistent with the initial definitions:
$
  \mathbb{E}_{\mu\nu} = \bm{n}(\partial_\mu c_\nu - \partial_\nu c_\mu) ,
$
$
  \mathbb{H}_{\mu\nu}  = - g \mathbb{B}_\mu \times \mathbb{B}_\nu .
$
Therefore, the dimension two composite operators $\mathbb{B}_\mu^2$ and $\sqrt{\mathbb{H}_{\mu\nu}^2}$, and the dimension four operators  $\mathbb{H}_{\mu\nu}^2$ are gauge invariant under the local U(1)$_{\rm II}$ gauge transformation \cite{Kondo04,KKMSS05}.


\end{document}